\documentclass[doublecol]{epl2} 
\usepackage{amsmath}
\usepackage{graphicx}
\usepackage{multimedia}
\usepackage{color}
\usepackage{bbold}
\usepackage{bm}

\allowdisplaybreaks

\newcommand{\mb}{\mathbf}
\newcommand{\up}{\uparrow}
\newcommand{\down}{\downarrow}

\newcommand{\comment}[1]{}

\title{Andreev spectrum of a Josephson junction with spin-split superconductors}
\shorttitle{Josephson junction with spin-split superconductors} %Insert here a short version of the title if it exceeds 70 characters

\author{B. Bujnowski\inst{1} \and D. Bercioux\inst{1,2} \and F. Konschelle\inst{3} \and J. Cayssol\inst{4} \and F. S. Bergeret\inst{1,3}}
\shortauthor{B. Bujnowski \etal}

\institute{                    
  \inst{1} Donostia International Physics Center (DIPC) - Manuel de Lardizabal 5, E-20018 San Sebasti\'{a}n, Spain\\
  \inst{2} IKERBASQUE, Basque Foundation of Science - 48011 Bilbao, Basque Country, Spain\\
  \inst{3} Centro de F\'isica de Materiales (CFM-MPC) Centro Mixto CSIC-UPV/EHU - 20018 Donostia-San Sebastian, Basque Country, Spain\\
  \inst{4} LOMA (UMR-5798), CNRS and Universit\'e Bordeaux, F-33045 Talence, France \\
}

\abstract{
The Andreev bound states and charge transport in a Josephson junction between two superconductors 
with  intrinsic exchange fields are studied. 
We find that for a parallel configuration of the exchange 
fields in the superconductors the discrete spectrum consists of two pairs of spin-split states. 
The Josephson current in this case is mainly carried by  bound states. In contrast, for the antiparallel
configuration we find that there is no spin-splitting of the bound states and that for phase differences smaller 
than certain critical value there are no bound states at all. Hence the supercurrent is only  carried by states in 
the continuous part of the spectrum.  Our predictions can be tested by performing a tunneling 
spectroscopy of a weak link between two spin-split superconductors.}

\begin{document}

\maketitle

\section{Introduction}
Superconductors with spin-split density of states have attracted particular interest since the pioneering works of Tedrow and Merservey, in which Zeeman splitting in superconductors was used to determine the spin-polarization 
of ferromagnetic metals \cite{PhysRevLett.26.192,meservey1994spin}.
Such spin-splitting can be achieved either by applying an external magnetic field or in thin superconducting films in contact with ferromagnetic insulators (FI) at zero field \cite{PhysRevLett.56.1746,Hao:1990}.
The spin-split density of states found in superconducting 
films originates from the exchange interaction between the conduction electrons of the superconductor
and the large localised magnetic moments of the FI \cite{PhysRevB.38.8823}.
In order to obtain large spin-splittings, the use of FIs has the advantage of avoiding the application of high magnetic fields.
The spectrum of  a conventional superconductor in this case shows two
BCS-like densities 
of states shifted by the energy $2h$, where $h$ is the effective exchange field induced in the superconductor film.  Here we denote 
them  as spin-split superconductors (SS).

There has been a resurgence of interest in SS because of several theoretical studies proposing them as absolute spin-valves
\cite{huertas2002absolute}, heat-valves \cite{Giazotto:2013ei} and thermoelectric elements \cite{MacHon2013,Ozaeta2014a,Giazotto:2014a}. Moreover, superconducting heterostructures  with  spin-splitting fields  have attracted the interest from theorists and experimentalists in the last years, mainly motivated by the possible detection of Majorana fermions \cite{PhysRevLett.104.040502,Science.349.6084,Nature531.7593,NaturePhysicsSzombati2016}
and elaboration of complex S-FI heterostructures \cite{PhysRevLett.108.127002,PhysRevLett.104.137002,PhysRevLett116.077001.2016}, where S denotes a BCS superconducting lead.

One striking effect in such structures is the enhancement of the critical Josephson current in a FI-S/I/FI-S junction by increasing the amplitude of 
the spin-splitting field\cite{bergeret2001enhancement,chtchelkatchev2002josephson,blanter2004}. Here I denotes an insulating tunneling barrier. This phenomenon has been demonstrated experimentally in Ref.\cite{robinson2010enhanced}. 

In order to understand the supercurrent in ballistic Josephson junctions it is important to analyze the spectral properties of these weak links~\cite{Holmqvist:2012,Stadler:2013}. In
a short ballistic superconductor-normal metal-superconductor (S/N/S) junction with equal gaps and at low temperatures, tunneling through Andreev bound states (ABSs) is the dominant contribution to the Josephson current \cite{Beenakker}.
The dependence of the ABSs on the phase difference between the superconducting banks in SS/I/SS junctions remains unexplored so far.

In this letter we investigate in detail a single-channel Josephson weak link connecting two spin-split (SS) superconducting leads. We focus on the dependence of sub-gap states on the superconducting phase difference across a ballistic SS/I/SS 
junction with a tunneling barrier of arbitrary strength. We extend the results \cite{bergeret2001enhancement,chtchelkatchev2002josephson,blanter2004}  
by demonstrating that any deviation from the case of equal exchange fields leads to the complete dissapearance of the ABSs in a finite range of the superconducting phase bias $\phi$ defined by a critical phase $\phi_\text{C}$ such that $|\phi|<\phi_C$. 
This phenomenon originates in the spin dependent asymmetry of the gaps in the left and right SS electrodes. As a consequence, within these interval the Josephson current is carried exclusively by states in the continuous part of the spectrum. 
The  value of $\phi_\text{C}$ does not depend on the transmissivity of the junction and hence it is robust against imperfections.

\section{Theory}
%
%
%%%%%%%%%
\begin{figure}
   \begin{center} 
   \includegraphics[width=\columnwidth]{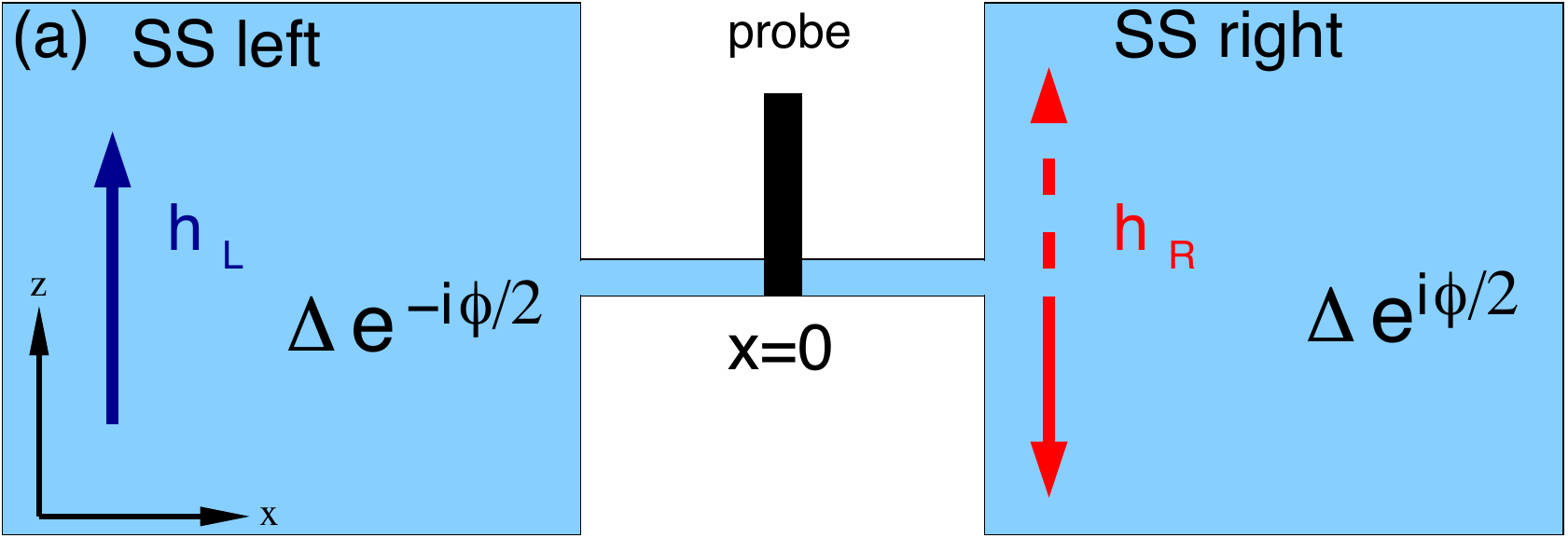}
   \includegraphics[width=\columnwidth]{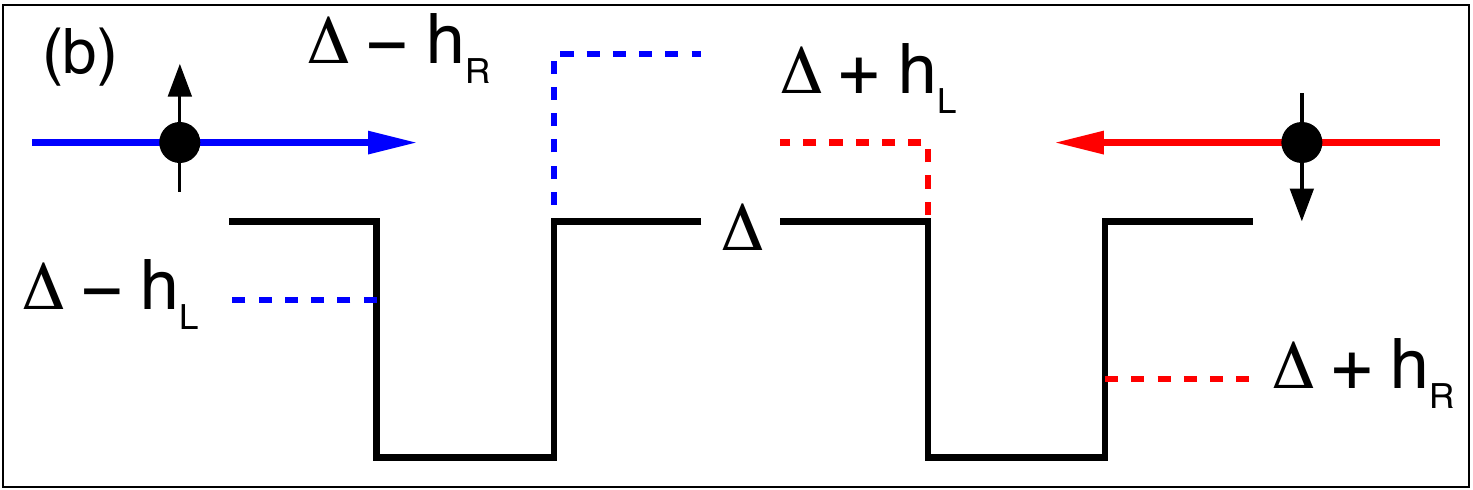}
   \caption{(a): Schematic diagram of the junction. Two SS electrodes  with intrinsic exchange fields $h_\text{L},\;h_\text{R}$ separated by a ballistic weak link.  A tunneling probe is situated at $x=0$.
   (b)~Sketch of the effective gaps for spin up/down electron when the left/right exchange fields $h_\text{L}/h_\text{R}$ are configured so that $-h_R>h_L>0$.}\label{junction}
   \end{center}
\end{figure}
%%%%%%%%%
%
%

We consider a Josephson junction consisting of two bulk SSs connected by a ballistic weak link [see Fig.~\ref{junction}(a)].
We model the weak link as a $\delta$-function scattering potential with strength $U$.
 The corresponding  Bogoliubov-de Gennes equation for quasiparticle states with
energy $E$ reads
\begin{align}\label{BDG}
\begin{pmatrix}
\hat{\mathcal{H}}_0(\mb{r}) & \hat{\Delta}(\mb{r})\\
\hat{\Delta}^\dagger(\mb{r}) & -\hat{\mathcal{H}}^\text{T}_0(\mb{r})
\end{pmatrix}\Psi(\mb{r})=E\Psi(\mb{r}),
\end{align}
where
%\begin{align}
%\hat{\mathcal{H}}_0(\mb{r})=&\left[-\frac{\hbar^2}{2m}\nabla_\mb{r}^2-\mu\right]
%\mathbb{1}+\delta(x)\cdot U\\\nonumber
%&-\hat{\bm{\sigma}}\cdot\left[\Theta(-x)\mb{h}_\text{L}+\Theta(x)\mb{h}_\text{R}\right] .
%\end{align}

\begin{align}
\hat{\mathcal{H}}_0(\mb{r})=&-\frac{\hbar^2}{2m}\nabla_\mb{r}^2-\mu+U\delta(x)\\\nonumber
&-\left[\Theta(-x)h_\text{L}+\Theta(x)h_\text{R}\right]\hat{\sigma}_z \, ,
\end{align}
and
\begin{align}\label{gap}
\hat{\Delta}(\mb{r})=\text{i}\hat{\sigma}_y\Delta[\text{e}^{-\text{i}\phi/2}\Theta(-x)+\Theta(x)\text{e}^{\text{i}\phi/2}] \, .
\end{align}
Here, $\hat{\sigma}_i$ are the Pauli matrices describing the spin degree-of-freedom. The temperature-dependent gap is modeled by the interpolation formula $\Delta=\Delta(T)\cong\Delta_0\tanh(1.74\sqrt{(T_\text{C}/T)-1})$, where $T_\text{C}$ is the critical temperature for superconductivity. In Eq.~\eqref{gap}, $\phi$ is the phase difference between the order parameters of the superconductors,  $\Theta(x)$ is
the Heaviside step function, and $\delta(x)$ is the Dirac delta function. 
We assume weak exchange fields so that the Clogston-Chandrasekhar criterion, $|h_\text{L,R}|< \Delta_0/ \sqrt{2}$, is fulfilled, where $\Delta_0$ is the BCS gap at zero temperature and zero exchange field \cite{PhysRevLett.9.266,Chandrasekhar}.  
We restrict ourselves to symmetric electrodes (in the absence of exchange fields) with equal gap magnitudes, chemical potentials and effective masses on both sides of the junction. The only asymmetry originates from the exchange fields in the left $(\text{L})$, right $(\text{R})$ superconducting leads, which are assumed to be collinear, though with arbitrary values $h_\text{L}$ and $h_\text{R}$. In this case the boundstate spectra can be obtained analytically.

We solve Eq.~\eqref{BDG} separately in the L and R region.
In the bulk SS we obtain plane-wave solutions with 
$\psi^\nu_{\bm{k},e(h),\sigma}(\bm{r})=\phi^\nu_{e(h),\sigma}\text{e}^{\text{i}\bm{k}^{e,(h)}_\sigma\bm{r}}$ for
electron-like (hole-like) quasiparticles with spin
$\sigma$. The spinors are 
\begin{subequations}
\begin{align}
\phi^\nu_{e,\up}&=(u^{\nu}_{\up}\text{e}^{\text{i}\phi_\nu},0,0,v^{\nu}_{\up})^\text{T},\\
\phi^\nu_{h,\up}&=(v^{\nu}_{\up}\text{e}^{\text{i}\phi_\nu},0,0,u^{\nu}_{\up})^\text{T},\\
\phi^\nu_{e,\down}&=(0,-u^{\nu}_{\down}\text{e}^{\text{i}\phi_\nu},v^{\nu}_{\down},0)^\text{T},\\
\phi^\nu_{h,\down}&=(0,-v^{\nu}_{\down}\text{e}^{\text{i}\phi_\nu},u^{\nu}_{\down},0)^\text{T}\,.
\end{align}
\end{subequations}
Here we have introduced the coherence factors $u^\nu_{\sigma}=\sqrt{(E^\nu_\sigma+\Omega^\nu_\sigma)/2E^\nu_\sigma}$,
$v^\nu_{\sigma}=\sqrt{(E^\nu_\sigma-\Omega^\nu_\sigma)/2E^\nu_\sigma}$, where
$\Omega^\nu_\sigma=\sqrt{{E^\nu_\sigma}^2-\Delta^2}$ and $E^\nu_\sigma=E+\sigma h_\nu$, $(\nu=\text{L},\text{R})$.
%
%
%%%%%%%%%%
\begin{figure*}
  \begin{center}\includegraphics[clip,width=2\columnwidth]{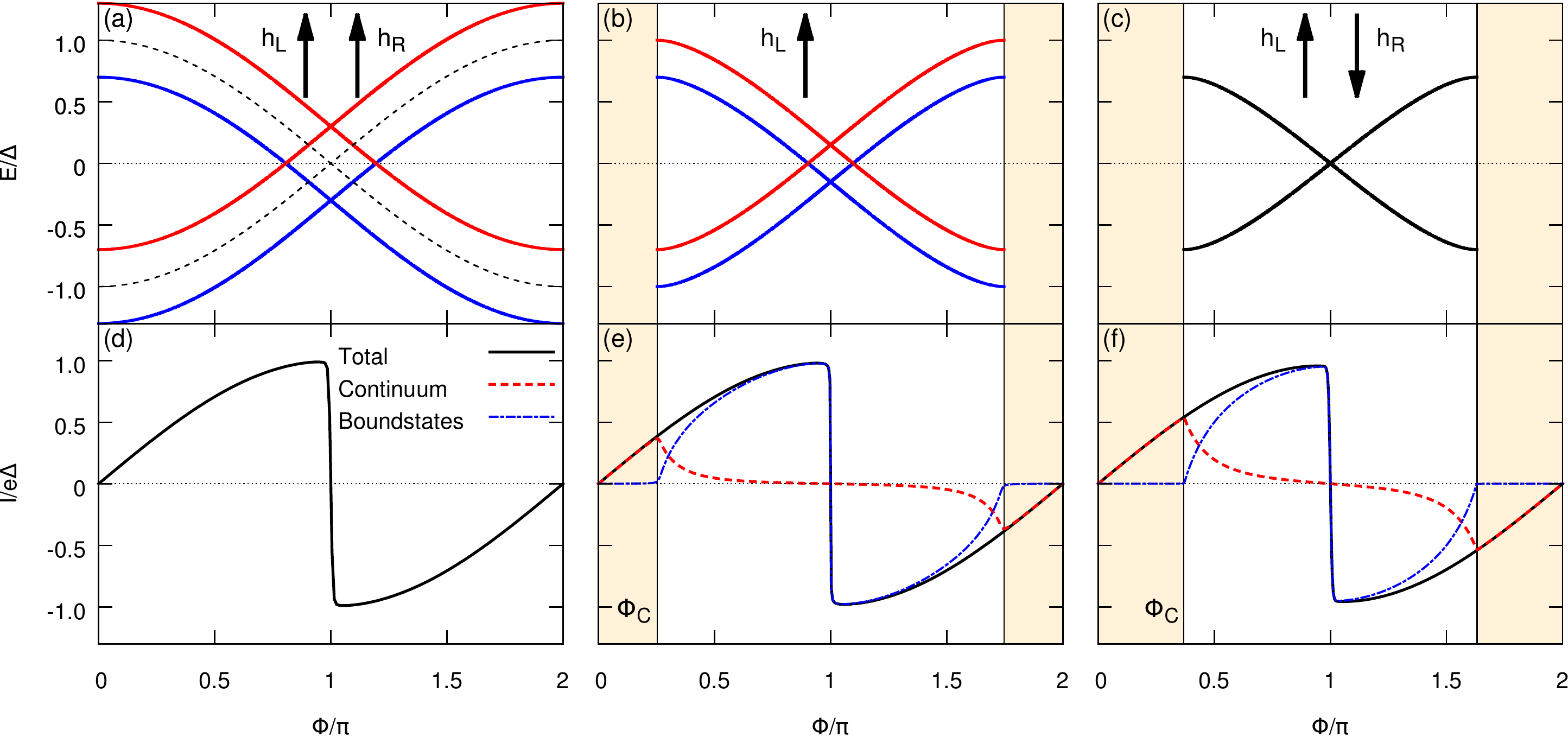}
  \end{center}
   \caption{Top panels (a) - (c) show Andreev bound state energies for (a) non-magnetic case (black dashed line) and parallel orientation of the exchange fields ($h_\text{L}=h_\text{R}=0.3\Delta$), (b) one side of the junction with zero 
   exchange field ($h_\text{L}=0.3\Delta, h_\text{R}=0$) and (c) anti-parallel orientation of the exchange fields ($h_\text{L}=-h_\text{R}=0.3\Delta$). The coloured regions correspond to the
   interval $\left|\phi\right|<\phi_\text{C}$ where there is no formation of Andreev bound states. The panels (d)-(f) show the corresponding current phase relationships.
   Where applicable we separated the continuum and bound state contribution to the total current (dashed red and dash-dotted blue lines). All plots are for $\tau=1$ and the current versus phase relationships
   are calculated at $T/T_\text{C}=0.01$.} \label{BS_multi}
\end{figure*}
%%%%%%%%%%%
%
%

We use these piecewise solutions to construct the
wave function ansatz for a spin-$\sigma$ electron-like quasiparticle
incident from the left SS with wavevector $\bm{k}^e_{\sigma}$.
In the following we consider a narrow wire constriction, and provide the corresponding single channel calculations. 
%We only consider trajectories perpendicular to the interface as the dependence of the angle of incidence
% is not important in the case of an isotropic gap and the chosen setup of a thin wire that can be well approximated
% by a 1D geometry. 
Therefore, the wave function ansatz reads:
\begin{align}\label{elinj}
&\Psi_{e,\sigma}(\bm{r})=\notag\\ &\Theta(-x)\Big\{\psi^\text{L}_{\bm{k},e,\sigma}+\sum_{\sigma^\prime=\up\down}
\big[a^e_
{\sigma\sigma^\prime}\psi_{\bm{k},h,\sigma^\prime}^\text{L}
+b^e_{\sigma\sigma^\prime}\psi^\text{L}_{-\bm{k},e,\sigma^\prime}\big]\Big\}\notag\\
&+\Theta(x)\Big\{\sum_{\sigma^\prime=\up\down}[c^e_{\sigma\sigma^\prime}\psi^\text{R}_{\bm{k},e,\sigma^\prime}+
d^e_{\sigma\sigma^\prime}\psi^\text{R}_{-\bm{k},h,\sigma^\prime}]\Big\}.
\end{align}
For $x<0$, Eq.~\eqref{elinj} describes the superposition of an incident electron-like quasiparticle with an Andreev reflected hole-like quasiparticle (with amplitude $a^e_{\sigma\sigma^\prime}$) 
and a reflected electron-like quasiparticle (with amplitude $b^e_{\sigma\sigma^\prime}$). For $x>0$, electron-like and hole-like quasiparticles are transmitted 
with
probability amplitudes $c^e_{\sigma\sigma^\prime}$ and 
$d^e_{\sigma\sigma^\prime}$, respectively. The ansatz for an
incident hole-like spin $\sigma$ quasiparticle $\Psi_{h,\sigma}(\bm{r})$ is analogous.
The probability amplitudes for this case are distinguished by the superscript
$h$, 
\emph{e.g.}, $a^h_{\sigma\sigma^\prime}$, $b^h_{\sigma\sigma^\prime}$ etc. We work within the so-called Andreev
approximation by assuming that $\mu\gg \max{(E,\Delta,|h_\nu|)}$, 
so that the electron and hole quasiparticle wavevectors can be regarded as approximately
equal in magnitude, $k^e_{\sigma}\approx k^h_{\sigma}\approx k_\text{F}$, where $k_\text{F}$ is the Fermi momentum in the normal state.

The probability amplitudes in Eq. (\ref{elinj}) for the various processes are calculated requiring the continuity of the wave function and  a finite jump of the derivative at the interface.
In particular  the Andreev reflection amplitudes~\cite{Andreev} read
\begin{align}\label{ABScoeff}
 a^e_{\sigma\sigma}=\frac{\Delta\left(E^\text{L}_\sigma\cos\phi-E^\text{R}_\sigma\right)+\text{i}\Delta\sin\phi\Omega^\text{L}_\sigma}{E^\text{L}_\sigma E^\text{R}_\sigma
 -\Delta^2\cos\phi+(1+2Z^2)\Omega^\text{L}_\sigma\Omega_\sigma^\text{R}} \, ,
\end{align}
and $a^h_{\sigma\sigma}(\phi)=a^e_{\sigma\sigma}(-\phi)$ where we introduced the dimensionless strength of the scattering potential $Z=2mU/k_\text{F}\hbar^2$ \cite{PhysRevB.25.4515}. The parameter  $Z$
is related to the transmission $\tau$ of the barrier as $\tau=1/(1+Z^2)$.

%The Andreev coefficients \eqref{ABScoeff} carry the 
%full information about the ABS in the system~\cite{Andreev}.

The complete Green's function of the junction is built from the scattering solutions Eq.~\eqref{elinj}~\cite{McMillan}. The retarded Green's function reads:
\begin{align}\label{GreensFK1}
&G^r(x,x',E)=\!\!\sum_\sigma\mathcal{A}_\sigma^\nu \left\{
 \left[a_{\sigma\sigma}^e \text{e}^{\text{i}k_\text{F}(x-x')}+a_{\sigma\sigma}^h \text{e}^{-\text{i}k_\text{F}(x-x')}\right]\times\right. \nonumber\\
& \left. 
\left(\begin{smallmatrix}
  u^\nu_\sigma v^\nu_{\bar{\sigma}} & {v^\nu_{\bar{\sigma}}}^2 \text{e}^{\text{i}\phi_\nu}\\
  {u^\nu_\sigma}^2 \text{e}^{-\text{i}\phi_\nu} & u^\nu_\sigma v^\nu_{\bar{\sigma}}
\end{smallmatrix}\right)
%%%%%%%%%%%%%%%%%%%%%
  +\right.\nonumber \\
&  \left.\left[\text{e}^{\text{i}k_\text{F}|x-x'|}+b_{\sigma\sigma}^e \text{e}^{-\text{i}k_\text{F}(x+x')} \right]
\left(\begin{smallmatrix}
  {u^\nu_\sigma}^2 & u^\nu_\sigma v^\nu_{\bar{\sigma}} \text{e}^{\text{i}\phi_\nu}\\
  u^\nu_\sigma v^\nu_{\bar{\sigma}}\text{e}^{-\text{i}\phi_\nu} & {v^\nu_{\bar{\sigma}}}^2
\end{smallmatrix}\right)
+ \right.\nonumber \\
&\left.\left[\text{e}^{-\text{i}k_\text{F}|x-x'|}+b_{\sigma\sigma}^h \text{e}^{\text{i}k_\text{F}(x+x')}\right]\left(\begin{smallmatrix}
  {v^\nu_{\bar{\sigma}}}^2 & u^\nu_\sigma v^\nu_{\bar{\sigma}} \text{e}^{\text{i}\phi_\nu}\\
  u^\nu_\sigma v^\nu_{\bar{\sigma}}\text{e}^{-\text{i}\phi_\nu} & {u^\nu_\sigma}^2
\end{smallmatrix}\right)\right\} ,
\end{align}
where $\mathcal{A}_\sigma^\nu=-\frac{\text{i}mE^\nu_\sigma}{\hbar^2k_\text{F}\Omega^\nu_\sigma}$.
This Green's function carries the complete information about the system and allows  the computation of the phase dependent local density of states (LDOS) and the Josephson current \cite{Beenakker,BeenakkerBrouwer}. The poles of the Green's function Eq.~\eqref{GreensFK1} give direct access to the whole spectrum of the Josephson junction: the discrete Andreev bound states coincide with poles of the Andreev reflection coefficients Eq.~\eqref{ABScoeff}, while the branch cuts of Eq.~\eqref{ABScoeff} provide the continuum part of the spectrum.

The LDOS at the tunneling barrier can be related to the $(1,1)$ component of the 
retarded Green's function Eq.~\eqref{GreensFK1} \cite{Kasiwaya_Tanaka}, using the formula $\rho(E,x)=\sum_\sigma\rho_\sigma(E,x)= -\lim_{x'\rightarrow x}\frac{1}{\pi}\text{Im}[G^r_{11}(x,x',E)]$. In our case, the spin-resolved LDOS reads
%
%
%%%%%%%%
\begin{align}\label{LDOS}
 \rho_\sigma(E)=\frac{m}{\pi\hbar^2k_\text{F}}\text{Re}\left[\left(\frac{2E^\nu_\sigma+(a^e_{\sigma\sigma}+a^h_{\sigma\sigma})\Delta}{2\Omega^\nu_\sigma}\right)\right]  \, ,
\end{align}
%%%%%%%%%
%
%
where the atomic scale oscillations of $\rho(E)$ are assumed to be averaged out\cite{PhysRevB.53.9371}.

In addition to the energy spectrum, we are also interested in the Josephson current through the junction
%
%
%%%%%%%%%%%%%
\begin{align}\label{IPHI}
&I = \frac{\text{i}e\hbar}{8\pi m}\int_{-\infty}^\infty dE \tanh\left(\frac{E}{2 k_\text{B} T}\right)\times \nonumber\\& \lim_{x'\rightarrow
  x=0^+}\!\!\!\left(\!\frac{\partial}{\partial{x}} -
\frac{\partial}{\partial{x'}}\!\right)\text{Tr}\left[G^{r}(x,x',E)-G^{a}(x,x',E)\right] ,
\end{align}
%%%%%%%%%%%%
%
%
where $k_\text{B}$ is the Boltzmann constant and $\text{Tr}[\ldots]$ is the trace in Nambu-spin space and $G^{r/a}$ is the retarded/advanced Green's function, using the real time representation of the Furusaki-Tsukada formula \cite{Furusaki,Furusaki:a}.
%
%
%%%%%%%%%%%%
\begin{figure}[!b]
\begin{center}
\includegraphics[clip,width=\columnwidth]{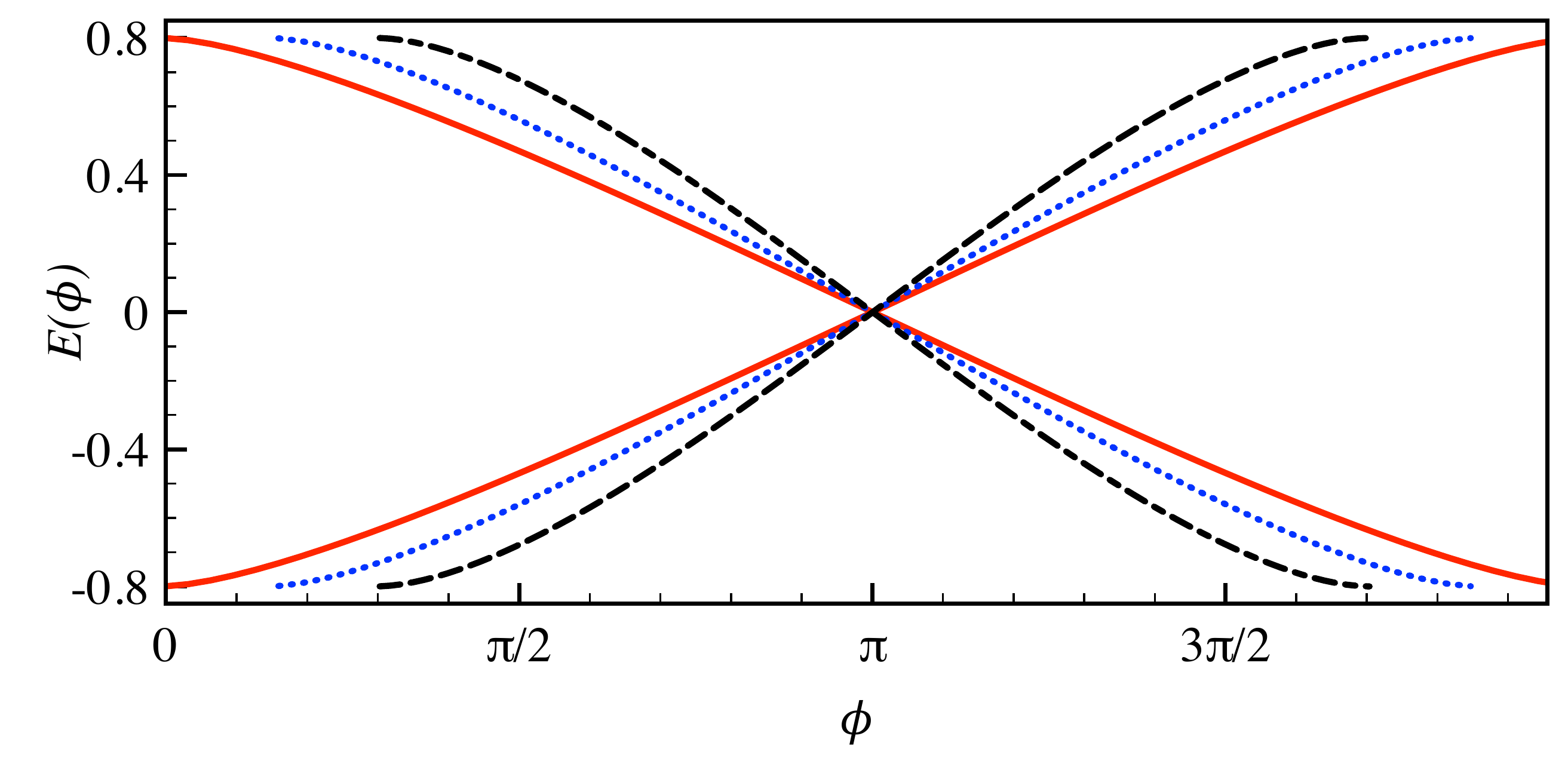}
\caption{\label{Length} Discrete part of the spectrum of a SS/N/SS junction as a function of the phase across the junction in the anti-parallel configuration ($h_\text{L}=-h_\text{R}=0.2$) for three different lengths of the junction: $L=0$ dashed-black line, $L=0.3\xi$ dotted-blue line and $L=0.6\xi$ solid-red line. }
\end{center}
\end{figure}
%%%%%%%%%%%%
%
%

\section{Results: Andreev bound states}
We start by analyzing the results for the discrete Andreev bound state (ABS) spectrum of a short junction with a perfect transmission coefficient ($\tau=1$), thereby recovering the well-known
phase dependence of the ABS energy in a short S/N/S junction without spin-splitting fields (Fig.~\ref{BS_multi}(a), black-dashed line).
In  the case of parallel exchange fields equal in magnitude ($h_\text{L}=h_\text{R}$) we find a splitting of the ABS energy-phase relationship of magnitude $|h_\text{L}+h_\text{R}|$ between
spin-up and spin-down quasiparticles (Fig.~\ref{BS_multi}(a), red and blue solid-lines). 

By lowering the value of  one  of the exchange fields while keeping the other fixed, the ABSs disappear within finite intervals of the phase difference $\phi$ [Figs.~\ref{BS_multi}(b), \ref{BS_multi}(c)]. Moreover this behaviour is independent of the transmission of the barrier. The minimal phase difference $\phi_\text{C}=\arccos(1-|h_\text{L}-h_\text{R}|/ \Delta)$ for which bound-states exist depends only on the difference between the exchange fields. In short, the ABSs are found only in the interval  $\phi\in[\phi_\text{C},2\pi-\phi_\text{C}]$. At the same time we observe a reduction of the gap.

One can provide a physical interpretation for the reduction of the gap and disappearance of ABS for some phase ranges: For illustration we consider a spin-up quasi-particle with positive energy coming from 
the left electrode [c.f. Fig.~\ref{junction}(b)], in the parameter regime with $h_\text{L}>0>h_\text{R}$, $|h_\text{R}|>|h_\text{L}|$. This quasi-particle encounters a reduced gap in the left SS of magnitude $\Delta-h_\text{L}$  and an enhanced  gap of magnitude $\Delta-h_\text{R}$ in the right SS. If the quasi-particle 
energy is higher than the energy of the left gap and lower than the right one, it can be Andreev reflected only at the right SS and ABSs can not be formed. The process for a spin-down quasi-particle
incoming from the right electrode is analogous. The same picture applies for quasi-particles with energies $E<0$ and can be modified to any case of collinear orientation of the exchange fields. 
This scenario is very similar to the case of a junction between two superconductors with gaps different in magnitude~\cite{Chang:1993}, where the existence of the ABSs was shown to be set by the smaller gap, but was completely spin independent. In the case of SSs leads, the distinct exchange fields induce the asymmetry between the gaps, which is different for spin up or spin down quasi-particles [Fig.~\ref{junction}(b)].

%In Ref.~\cite{Chang:1993} the appearance of the critical phase was explained
%The appearance of the critical phase is also related to the presence of an asymmetry between gaps~\cite{Chang:1993}.
%
%
%%%%%%%%%%
\begin{figure}[!hb]
\includegraphics[width=\columnwidth]{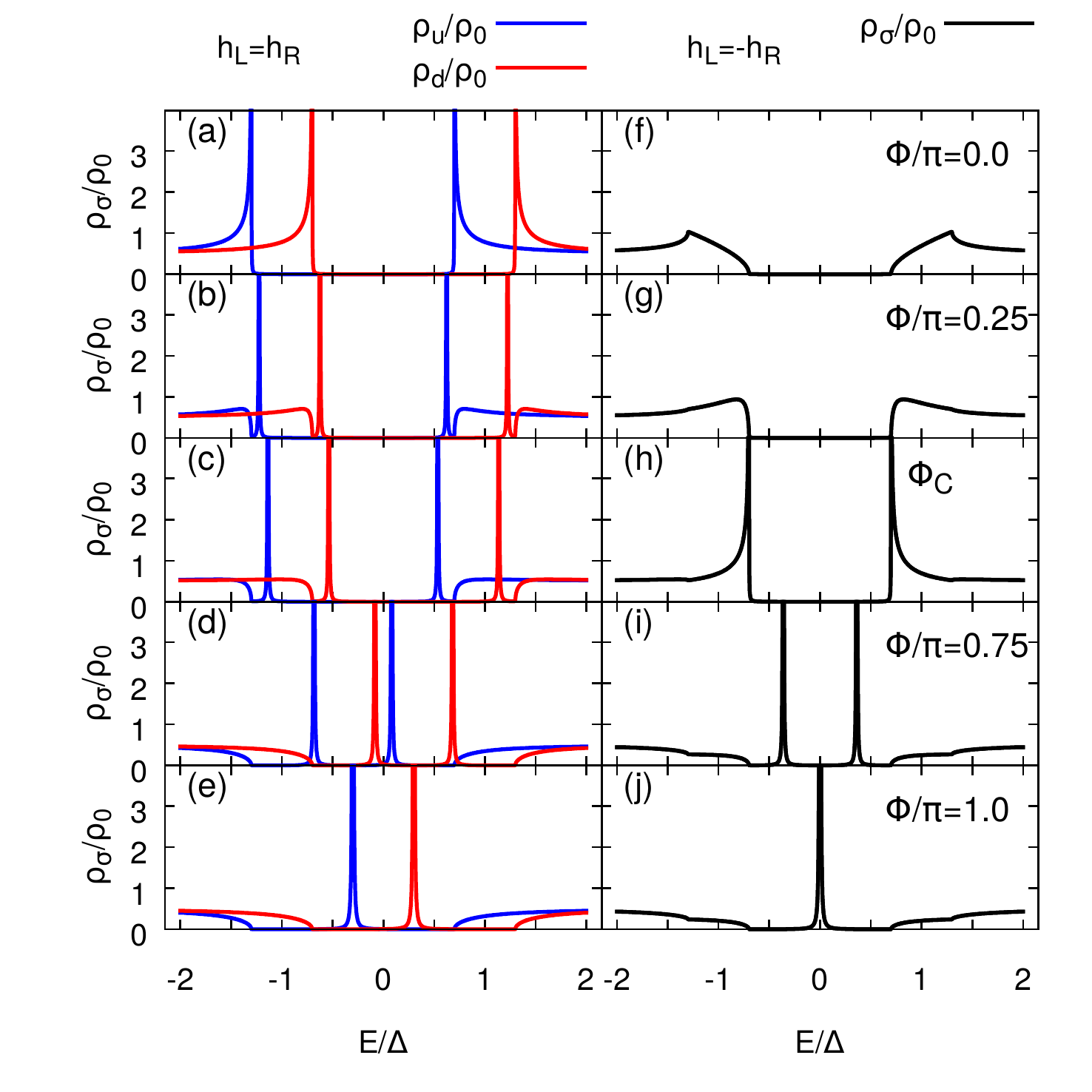} \caption{LDOS of the junction $\rho(E)$ divided by the normal state LDOS $\rho_0(E)$, for parallel (left column) and anti-parallel 
(right column) orientation of the exchange fields of magnitude $|h_{L,R}|/\Delta=0.3$. The phase difference gradually increases from the
   top panels where $\phi=0$ through $\phi_\text{C}$ to the bottom panels with $\phi=\pi$. All plots are for $\tau=1.0$.}\label{LDOS_comp}
\end{figure}
%%%%%%%%%
%
%
The above results where obtained for short weak links, \emph{i.e.} for $L \ll \hbar v_\text{F}/\Delta$, where $L$ is the length of the normal region separating the two superconducting leads, and $v_\text{F}$ is the Fermi velocity. We now discuss whether the previous picture (spin dependent reduced gaps and disappearance of ABS) holds for longer junctions. For arbitrary lengths of the junction and a fully transparent link ($U=0$), the critical phase can be obtained by analyzing the Bohr-Sommerfeld quantization condition for the SS/N/SS junction, where we assume no magnetic field in the normal region~\cite{Zagoskin}:
%
%
%%%%%%%%%%%%
\begin{equation}\label{BohrSommerfeld}
2\frac{EL}{\hbar v_\text{F}} \pm\phi-\!\!\!\!\sum_{\nu=\{\text{L,R}\}}\!\!\arccos\left(\frac{E+\sigma h_\nu}{\Delta}\right)=2n\pi \,,
\end{equation}
%%%%%%%%%%%%
%
%
%
%
%%%%%%%%%%
\begin{figure*}
  \begin{center}\includegraphics[clip,width=2\columnwidth]{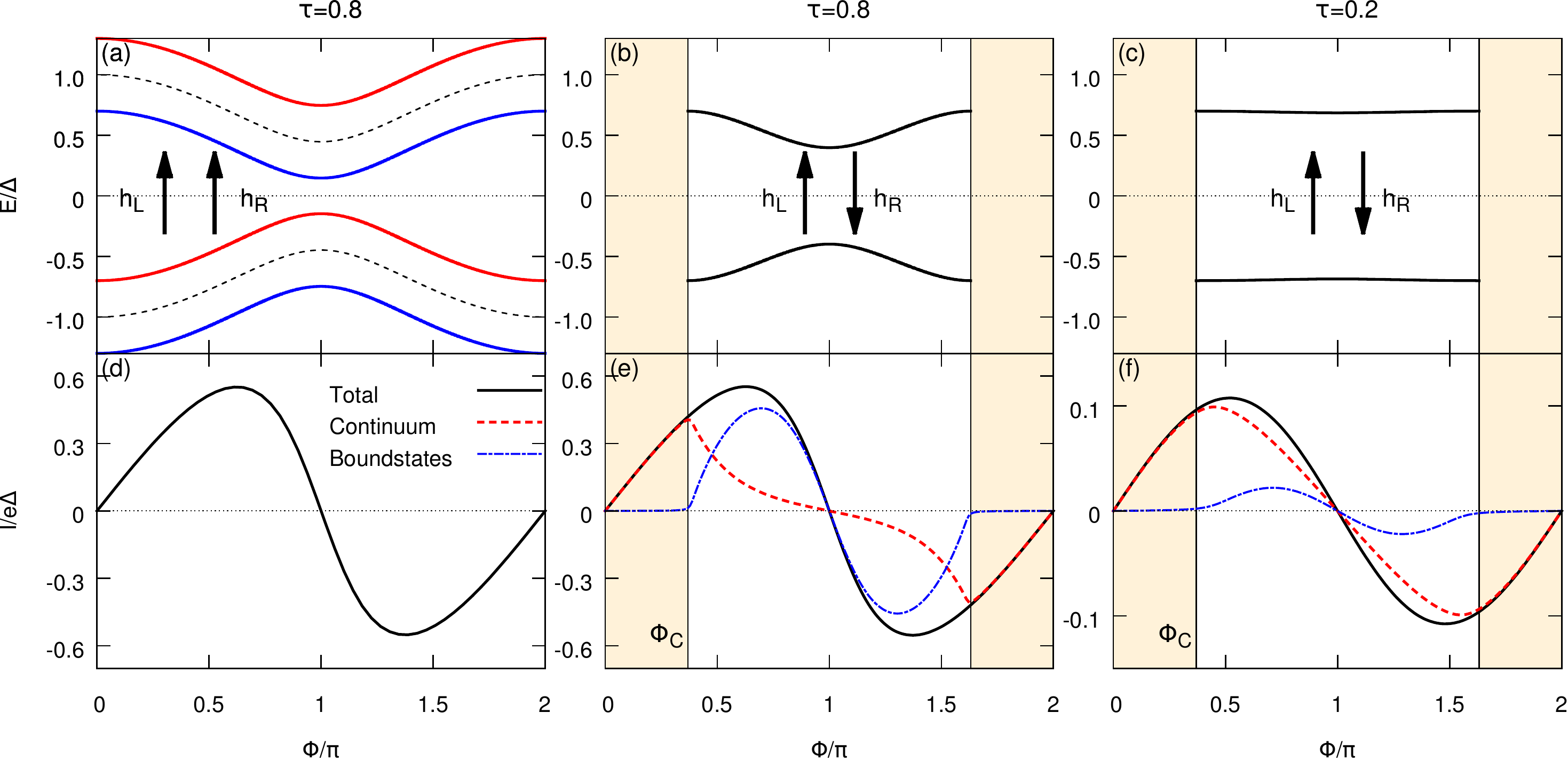}
   \end{center}
\caption{Andreev bound state energies for (a) non magnetic case (black dashed) and parallel orientation of the exchange fields ($h_\text{L}=h_\text{R}=0.3\Delta$), and (b)-(c) anti-parallel orientation of the exchange fields ($h_\text{L}=-h_\text{R}=0.3\Delta$). For panels (a),(b) $\tau=0.8$ and for (c) $\tau=0.2$. The coloured regions correspond to the intervals with no formation of Andreev bound states. Panels (d)-(f) show the corresponding current phase relationships separated into the continuum states (red dashed line) and bound states contribution (blue dash-dotted line) to the total current. All plots are for $T/T_\text{C}=0.01$.}\label{BS_multi_Z}
\end{figure*}
%%%%%%%%%%
%
%
with $n\in\mathbb{Z}$. Note that the spin-splitting of the gaps (being a bulk property of the SS leads) is independent of the length $L$. In the short junction limit
($L \ll \hbar v_\text{F}/\Delta$), one recovers the critical phase $\phi_\text{C}=\arccos(1-|h_\text{L}-h_\text{R}|/ \Delta)$ introduced earlier. 
From Eq.~\eqref{BohrSommerfeld} we can also infer the dependence of the critical phase on the length $L$ of the weak link: $\phi_\text{C}$ decreases as the length $L$ is increased, and $\phi_\text{C} \rightarrow 0$, for lengths exceeding the superconducting coherence length $\xi=\hbar v_\text{F}/\Delta$ (Fig.~\ref{Length}). Indeed in the long junction limit, even if the highest ABS merges into the continuum, there are other ABS (with lower energies) which are still defined for all values of the phase. Note that the study of ABSs associated with the Bohr-Sommerfeld quantization condition Eq.~\eqref{BohrSommerfeld} can be generalized to the case of a spin-active weak link using the formalism developed in Ref.~\cite{PhysRevLett.116.237002}.

\section{Results: local density of states} Direct insight about states for all phases can also be obtained by calculating the LDOS of the junction using Eq.~\eqref{LDOS}. 
In the parallel case and $\phi=0$, we obtain (as expected) the spectrum of a bulk SS with 
the two spin-split BCS densities of states with coherence peaks at $E=\pm(\Delta+\sigma h)$ [Fig.~\ref{LDOS_comp}(a)]. For a finite phase
difference between the SSs, spin-split ABSs appear. The peaks corresponding to hole- and electron-like quasiparticles with spin $\sigma$ are centered around $E=\sigma|h_\nu|$
[red (blue) lines in the left column in Fig.~\ref{LDOS_comp}] and merge at this energy when approaching $\phi=\pi$.

In the anti-parallel case and  $|\phi|<\phi_\text{C}$ [see Fig.~\ref{LDOS_comp}(f) and \ref{LDOS_comp}(g)] the spectrum deviates drastically from the BCS-like spectrum and no BCS coherent peaks are observed. At the critical value of the phase  $\phi_\text{C}$ these peaks appear at energies $\pm(\Delta-|h_\nu|)$. The two peaks corresponding to ABSs merge into a single peak at $\phi=\pi$ [Fig.~\ref{LDOS_comp}(j)].

For imperfect transmission ($\tau<1$) and parallel configuration of the exchange fields ($h_\text{R}=h_\text{L}=h$) [Fig.~\ref{BS_multi_Z}(a)] the energy difference between the spin polarized ABSs remains the same as in the $\tau=1$ case
In contrast, in the anti-parallel case there is no splitting of the ABSs. In both cases there are avoided energy crossings at $\phi=\pi$ due to finite backscattering. Noticeably, neither the spin-splitting
nor the critical phase $\phi_\text{C}$ are $\tau$-dependent.

\section{Results: current-phase relation}
To understand how the absence of ABSs influence the Josephson current in the non-parallel case for $\left|\phi\right|<\phi_\text{C}$, we numerically evaluate Eq.~\eqref{IPHI}. 
In the lower panels of Fig.~\ref{BS_multi} current phase relations are shown for different orientations of the exchange fields and 
 perfect transmission of the barrier ($\tau=1$). The current phase relations show the well-known  sawtooth shape Fig.~\ref{BS_multi}(d)-\ref{BS_multi}(f). 
 Lowering the transmission, the current phase relationships become sinusoidal and one recovers the usual current-phase relation of a tunneling junction, see Figs.~\ref{BS_multi_Z}(d)-\ref{BS_multi_Z}(f).
 We also verified the enhancement of the critical current with respect to the non-magnetic case 
by the presence of anti-parallel exchange fields in the low transmission limit\cite{bergeret2001enhancement,chtchelkatchev2002josephson}.

The total current is the sum of two contributions:  one originating from the ABS ($I_\text{ABS}$) and the other from states in the continuous spectrum ($I_\text{Cont}$). These  are shown in the lower panels of Figs.~\ref{BS_multi} and~\ref{BS_multi_Z}.
In the parallel configuration with identical exchange fields the Josephson current is carried exclusively by the ABSs. In contrast, if the exchange fields are different both $I_\text{ABS}$ and $I_\text{Cont}$ contribute to the current. The vanishing contribution from the discrete spectrum for $\left|\phi\right|<\phi_\text{C}$ is compensated by a finite  $I_\text{Cont}$ [see Figs.~\ref{BS_multi}(e) and \ref{BS_multi}(f) for $\tau=1$ and in Figs.~\ref{BS_multi_Z}(e) and \ref{BS_multi_Z}(f) for $\tau<1$].
In other words, current from tunneling through 
ABSs is only present for $\phi\not\in(-\phi_\text{C},\phi_\text{C})$ and gets reduced by lowering the transmission of the junction. High enough exchange interactions
and low transmission can lead to a current dominated by contributions from continuum states, as seen in Fig.~\ref{BS_multi_Z}(f). 
This is consistent with the results of Chtchelkatchev~\emph{et~al.}\cite{chtchelkatchev2002josephson} where the critical current is shown to be purely due to the states of the continuous spectrum 
in the case of high magnitudes of the anti-parallel exchange fields and sufficiently low transmissions.
\section{Conclusion}
We have presented a detailed study of the spectrum and current-phase relation of a Josephson junction consisting of a short weak link connecting two superconducting leads with a spin-split density of states. We have shown that for collinear orientations of the exchange fields, any deviation from the case of equal fields leads to finite intervals  of phases without Andreev bound states. 
These intervals are independent of the transmission of the junction and are characterized by a critical phase-difference $\phi_\text{C}=\arccos(1-|h_\text{L}-h_\text{R}|/\Delta)$ for which ABSs disappear by merging within the continuum. 

When the phase difference is in the range $\left|\phi\right|<\phi_\text{C}$, the Josephson current is therefore completely carried by states in the continuous part of the spectrum. Outside this range the current is a superposition of the contributions from the ABS and the continuous spectrum. For perfect transmission the current is mainly due to tunneling through the ABSs [Fig.~\ref{BS_multi}(e)], whereas for low transmission the current is totally due to excitations from the continuous part of the spectrum [Fig.~\ref{BS_multi_Z}(f)]. Hence changing the transmission of the junction allows to tune the origin of the current.

Our findings on the spectrum of SS/I/SS junctions  can be tested by
 tunneling spectroscopy of the ABS spectrum as in Ref.~\cite{nphys1811,nature2013}  by using 
 electrodes made of ferromagnetic insulator-superconducting bilayers, \emph{e.g.} EuS-Al~\cite{Hao:1990}, 
 coupled by a thin normal nanowire in a closed loop. 
 The full current-phase relation can be determined by  means of a tunneling probe (c.f. Fig.~\ref{junction})  in the middle of the N  wire.
\comment{ By means of a tunneling probe (c.f. Fig.~\ref{junction})  in the middle of the N  wire one can determine the full current-phase relation. }

\acknowledgments
We thank P.M.R. Brydon, V.N. Golovach and S. Kawabata for useful discussions. 
This  work  was supported by the  Spanish Ministerio de Economia y Competitividad (MINECO) through  Project No. FIS2014-55987-P and the Basque Departamento de Educacion, UPV/EHU  through  Project  IT-756-13, and by the (LTC) QuantumChemPhys.

\end{document}